\documentclass[aps,prb,showpacs,superscriptaddress,twocolumn]{revtex4-1}

\usepackage{amsmath,graphics}
\usepackage[next]{inputenc}
\usepackage[dvips]{epsfig}
\usepackage[colorlinks=true,citecolor=blue,linkcolor=blue]{hyperref}

\def\be{\begin{equation}}
\def\ee{\end{equation}}

\def\bi{\begin{itemize}}
\def\ei{\end{itemize}}
\def\bn{\begin{enumerate}}
\def\en{\end{enumerate}}
\def\bea{\begin{eqnarray}}
\def\eea{\end{eqnarray}}

\def\ba{\begin{array}}
\def\ea{\end{array}}
\def\bd{\begin{displaymath}}
\def\ed{\end{displaymath}}

\begin{document}
\title{Unusual magnetic phases in the strong interaction limit of two-dimensional topological band insulators in transition metal oxides}
\author{Mehdi Kargarian}
\email{kargarian@ph.utexas.edu}
\affiliation{Department of Physics, The University of Texas at Austin, Austin, TX 78712, USA}

\author{Abdollah Langari}
\affiliation{Department of Physics, Sharif University of Technology, Tehran 11155-9161, Iran}
\affiliation{Max-Planck-Institut f\"ur Physik Komplexer Systeme, 01187 Dresden, Germany}

\author{Gregory A. Fiete}
\affiliation{Department of Physics, The University of Texas at Austin, Austin, TX 78712, USA}
\begin{abstract}
The expected phenomenology of non-interacting topological band insulators (TBI) is now largely theoretically understood. However, the fate of TBIs in the presence of interactions remains an active area of research with novel, interaction-driven topological states possible, as well as new exotic magnetic states.  In this work we study the magnetic phases of an exchange Hamiltonian arising in the strong interaction limit of a Hubbard model on the honeycomb lattice whose non-interacting limit is a two-dimensional TBI recently proposed for the layered heavy transition metal oxide compound, (Li,Na)$_2$IrO$_3$. By a combination of analytical methods and exact diagonalization studies on finite size clusters, we map out the magnetic phase diagram of the model. We find that strong spin-orbit coupling can lead to a phase transition from an antiferromagnetic Ne\'el state to a spiral or stripy ordered state.  We also discuss the conditions under which a quantum spin liquid may appear in our model, and we compare our results with the different but related Kitaev-Heisenberg-$J_2$-$J_3$ model which has recently been studied in a similar context.

\end{abstract}
\date{\today}



\maketitle
\section{Introduction \label{intro}}
Topological band insulators (TBI) preserve time reversal symmetry, have a bulk band gap originating in strong spin-orbit coupling, and possess the physical signature of an odd number of gapless Dirac nodes at time-reversal invariant momenta on boundary.\cite{kane1:prl05,kane2:prl05,Bernevig:science06,Moore:prb07,Fu2:prb07,Fu:prb07,Teo:prb08} The spin-momentum locking originating from strong spin-orbit coupling in these systems and odd number of Dirac points results in gapless boundary states immune to Anderson localization.\cite{Schnyder:prb08}  Since the experimental discovery\cite{Konig:science07,Hsieh:nature2008} of these topological states of matter, there has been an explosion of both theoretical\cite{HasanKane:rmp10,Qi:rmp11,HasanMoore:Ann11,Moore:nature10} and experimental\cite{Hsieh:science09,Chen:science09,Xia:np09,Hsieh:prl09,Kuroda:prl10,Sato:prl10} works aimed at unraveling the fascinating properties of topological insulators.  Noninteracting topological insulators and superconductors can be classified in terms of topological invariants\cite{kane2:prl05,Moore:prb07,roy:prb09,ryu:NJP10} where the band topology fully characterizes the properties of the insulator and superconductor. What remains to be understood is the full effect of Coulomb interactions, including the possible magnetic phases which could arise from the nontrivial band topology in the limit of intermediate to strong interactions.  Almost all experimentally discovered topological insulators to date can be understood within a single-particle picture ({\it i.e.}, band theory) where the correlation effects are weak.\cite{hasan:arxiv10075111,wang:prb11R,lin:NJP11,wang:NJP11}  Besides the known topological insulators, there are scores of other materials whose noninteracting and weakly interacting limits have been predicted to be topological insulators.\cite{HasanKane:rmp10,Qi:rmp11,HasanMoore:Ann11,Hasan:arxiv11}

It is well known that the nontrivial band topology originates from relativistic spin-orbit coupling strong enough  to cause a band inversion at an odd number of time-reversal invariant momenta in the Brillouin zone.\cite{Bernevig:science06,Murakami:prb07} Hence, it is natural to expect that the materials with heavy elements with their large spin-orbit coupling may provide fertile ground in the search for topological insulators. Among them, the transition metal oxides with $5d$ elements such as $\mathrm{Ir}$ and $\mathrm{Os}$ attract attention as the $5d$ orbitals are spatially more extended, and therefore less correlated compared to the $3d$ and $4d$ orbitals. A few examples are the pyrochlore iridates, $\mathrm{A_2Ir_2O_7}$ (A is a rare earth element), the layered compounds $\mathrm{(Li,Na)_2IrO_3}$ and $\mathrm{Sr_2IrO_4}$, and the hyperkagome $\mathrm{Na_2Ir_4O_8}$. Naively, these materials are expected to be metallic with a partially filled transition metal ion shell and a relatively small effective Hubbard $U$ (on the order of 0.5-2.0 eV). However, since the intrinsic spin-orbit coupling is strong (on the order of 0.4-1.0 eV), the spin and orbital degrees of freedom are entangled\cite{Kim:science09} which can dramatically influence the band topology of the systems.\cite{shitade:prl09,Pesin:np10,Kargarian:prb11,yang:prb10}

The effect of electron interactions on band topology has been studied for variety of models.  A particularly well studied model is the famous Kane-Mele model\cite{kane1:prl05,kane2:prl05} with a Hubbard interaction added.\cite{wen:prb11,vaezi:prb12,rachel:prb10,Young:prb08,Hohenadler:prl11,Zheng:prb11,Wu:prb12,Hohenadler:prb12}  In the absence of spin-orbit coupling, the model reduces to the Hubbard model on the honeycomb lattice. Quantum Monte Carlo simulations were used to map out the phase diagram of the Hubbard model. Three phases were found: (i) a metallic phase, (ii) a quantum spin liquid (QSL), and (iii) an antiferromagnetic (AFM)  phase with increasing the strength of local Hubbard term.\cite{Meng:nature10} While the metallic phase is immediately converted into the quantum spin Hall state (QSH) upon the inclusion of a second-neighbor spin-orbit coupling, the QSL phase is stable over a small range of spin-orbit coupling.\cite{Hohenadler:prl11,Zheng:prb11,Wu:prb12,Hohenadler:prb12} Both QSH and QSL phases are unstable in the strong interacting limit where an in-plane antiferromagnetic ordering arises.\cite{Hohenadler:prl11,rachel:prb10,Soriano:prb10} In fact, recent QMC results suggests the QSL phase may not be present at all in the Kane-Mele-Hubbard model.\cite{Sorella:1207.1783}

Besides the two-dimensional Kane-Mele-Hubbard model, three dimensional systems have drawn attention.  Pyrochlore oxides with heavy transition elements have been studied using a strong spin-orbit coupling approach that splits the $t_{2g}$ manifold into a lower $j=3/2$ manifold and a higher $j=1/2$ manifold.  The latter acts effectively as a spin-1/2 degree of freedom.  When slave-particle approaches are applied, exotic topological Mott insulators with topologically protected gapless boundary spin excitations appear in a range of intermediate strength Hubbard interactions.\cite{Pesin:np10,Kargarian:prb11,Witczak-Krempa:prb10}  If time-reversal symmetry is broken, Weyl semi-metals may also appear in the pyrochlores\cite{Wan:prb11,Witczak-Krempa:prb12,Wan:prl12,Yang:prb11} and possibly also an axion insulator phase proximate to the Weyl semi-metal.\cite{Wan:prb11,Wan:prl12,Go:prl12}

In addition to the studies mentioned above, there are other works addressing the physics of interaction-generated spin-orbit coupling which could drive the system to a phase with nontrivial band topology. In this case, the topological order appears  via spontaneously generated complex hopping terms which mimic those of an intrinsic spin-orbit coupling. For two dimensional systems with quadratic band touching points in their non-interacting band structure,  the leading instability would be a quantum anomalous Hall (QAH) effect and/or a topological insulator which, respectively, break the time reversal symmetry and spin rotational symmetry.\cite{raghu:prl08,vishwanath:prb09,sun:prl09, wen:prb10,liuQin:prb10,Fiete:pe11}  Similar physics is also believed to possibly generate topological phases in transition metal oxide heterostructures derived from the much lighter 3$d$ elements.\cite{Ruegg:prb12,Ruegg11_2,Yang:prb11}   

In the strong coupling limit, spin-orbit coupling can also affect the magnetic phases of the transition metal oxides. In the absence of spin-orbit coupling and orbital degeneracy the strong coupling limit can often be adequately described in terms of a pure spin Hamiltonian of the Heisenberg form. This is believed to be the case in the insulating parent compound of cuprate superconductors, for example. However, an orbital degeneracy is often present in transition metal ions leading to a spin-orbital exchange interaction.\cite{khomski:spu82} In contrast to a single orbital model, the resulting exchange interaction could be highly anisotropic and frustrated.\cite{van:prl10,chern:pre11} The entangled spin and orbital states break the $\mathrm{SU(2)}$ symmetry of the magnetic Hamiltonian giving rise to realizations of exotic spin models such as the Kitaev\cite{Kitaev20032} or Heisenberg-Kitaev models\cite{Jackeli:prl09,Chaloupka:prl10} in transition metal oxides.

Recently, the layered perovskites $\mathrm{(Li,Na)_2IrO_3}$ have been suggested to host exotic phases. Temperature dependent electrical resistivity and magnetic measurements clearly indicate their insulating nature with enhanced magnetic correlations at low temperatures.\cite{singh:prb10} The insulating ground state is thought to be interaction-driven and magnetically ordered at low temperatures.\cite{jin:arxiv09} Although strong Coulomb interactions make the realization of topological band insulators unlikely, the intrinsic spin-orbit coupling does substantially modify the effective spin Hamiltonian: The Heisenberg-Kitaev model\cite{Chaloupka:prl10} has been proposed to explain the strong suppression of magnetic correlations due to the possible proximity to a quantum spin liquid phase ($T_N \approx 15 \mathrm{K}$ is much smaller than the Curie-Weiss temperature $\theta=-116$\cite{Reuther:prb11}), though recent x-ray magnetic scattering experiments suggest the system is magnetically zig-zag ordered.\cite{Liu:prb11} Subsequent works based on magnetic models might be able to describe this ordered phase.\cite{Kimchi:prbR11,Bhattacharjee:1108.1806,singh:prl12,Ruether:prb11,Jiang:prb11}  

In this work, we consider an alternative magnetic Hamiltonian that is obtained in the strong interaction limit of the model introduced in Ref.[\onlinecite{shitade:prl09}], which at the noninteracting level exhibits a two-dimensional $\mathrm{Z_2}$ topological insulator, and in the intermediate interaction regime may exhibit a novel interaction-driven topological insulator with a non-trivial ground state degeneracy and topologically protected {\em collective modes}.\cite{Ruegg:prl12}  In our derivation of the effective spin Hamiltonian, we explicitly include the second-neighbor real hopping (in addition to the complex hopping) predicted from band theory which then gives rise to an anisotropic exchange coupling, which fully breaks the $\mathrm{SU(2)}$ spin symmetry [see Eq.\eqref{H}].  By a combination of analytical and exact diagonalization studies we map out the phase diagram of the model. The full phase diagram is shown in Fig.\ref{PDJ2J3} where a variety of magnetic phases result from the interplay between spin-orbit coupling and correlations. We also discuss the partial relevance of the model to a magnetically ordered state discovered in layered $\mathrm{(Li,Na)_2IrO_3}$ as the ground state with stripy order still possess some degree of zig-zag ordering.

Our paper is organized as follows. In Sec.\ref{model} we introduce both the non-interacting and magnetic exchange Hamiltonians. In Sec.\ref{schwinger_sec} we use Schwinger Boson mean field theory (SBMFT) to address the ineffectiveness of the anisotropic exchange term in stabilizing a spin liquid phase. We then study possible classical magnetic phases in Sec.\ref{condicate_phases}, and in Sec.\ref{lanczos_sec} exact diagonalization is used to study the various magnetic phases and phase transitions between them. We further study the locations of critical points by use of fidelity in Sec.\ref{fidelity_sec}, and present our conclusions in Sec.\ref{conclusion}. Some details of the SBMFT are included in Appendix \ref{bosonH} and \ref{Hk}.     
\begin{figure}[t]
\begin{center}
\includegraphics[width=8cm]{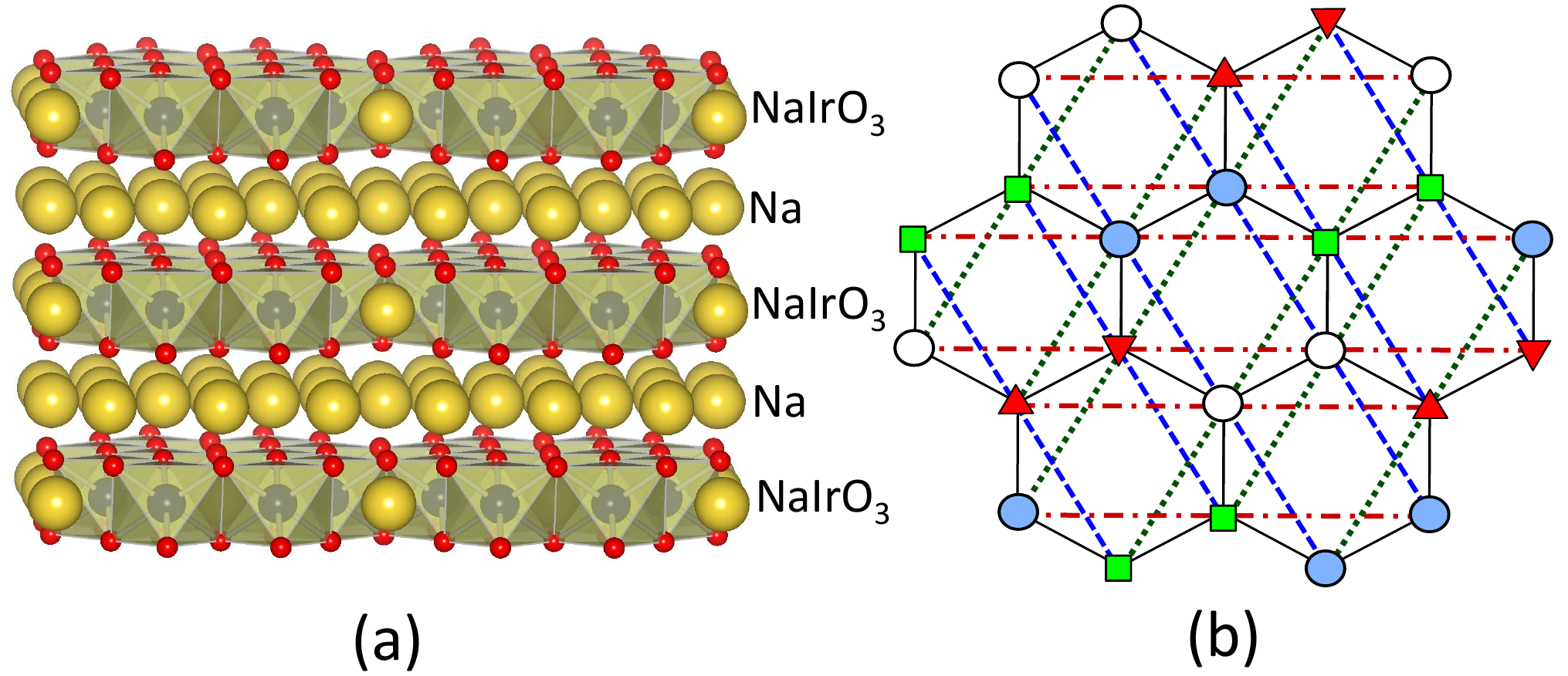}
\caption{(Color online) (a) Lattice representation of $\mathrm{Na_2IrO_3}$. The blue (grey), red (dark) and yellow (light) balls indicate Iridium, Oxygen and Sodium, respectively. The  Ir$^{+4}$ ions are located on the vertices of the honeycomb lattice of $\mathrm{NaIrO_3}$, stacked along c-axis.\cite{VESTA} (b) A top view of a honeycomb lattice. Solid lines stand nearest-neighbor (NN) coupling and red (dotted-dashed), green (dotted) and blue (dashed links) indicate the spin-dependent next-nearest-neighbor (NNN) hopping terms in Eq.\eqref{H0} involving the Pauli spin matricies $\sigma^{x}$, $\sigma^{y}$, and $\sigma^{z}$, respectively. The same NNN lines also stand for the anisotropic exchange couplings in Eq.\eqref{H}. Vertices shown by light and blue circles, squares and triangles indicate the local transformations on the triangular lattice which restores the $\mathrm{SU(2)}$ symmetry of third term in Eq.\eqref{H} [See text around Eq.\eqref{transformation}]. } \label{lattice}
\end{center}
\end{figure}

\section{model and methods\label{model}}
The layered oxides $\mathrm{(Li,Na)_2IrO_3}$ are composed of $\mathrm{NaIrO_3}$ layers stacked along the c-axis and separated by a layer of $\mathrm{Na}$ [see Fig.\ref{lattice}(a)] (and likewise for the Li-based material), where the transition metal ions $\mathrm{Ir^{+4}}(5d^{5})$ site on the vertices of a honeycomb lattice. The measured magnetic moment is  $\mu_{eff}=1.8$ $\mu_{B}$ verifying the description of the model in terms of local moments with $S_{eff}=1/2$.\cite{singh:prb10} The noninteracting model was argued, from a tight-binding fit to a density functional theory calculation, to be described by the following Hamiltonian,\cite{shitade:prl09}
\bea 
\label{H0} 
H_0=-t\sum_{<ij>\alpha}c^{\dag}_{i\alpha}c_{i\alpha}+\sum_{\ll ij\gg \gamma}\sum_{\alpha\beta}c^{\dag}_{i\alpha}t^{\gamma}_{\alpha\beta}c_{i\beta},
\eea 
where $c_{\alpha}^{\dag}(c_{\alpha})$ stands for the creation (annihilation) operator of an electron in a spin-orbital coupled pseudospin-1/2. The first term describes spin-independent nearest-neighbor hopping, while the second term describes second-neighbor spin-dependent hopping. The nearest neighbor hopping term gives a semi-metallic phase with Dirac nodes in the absence of second-neighbor hopping.  However, the second-neighbor hopping term, $t^{\gamma}=-t'\sigma^{0}+it''\sigma^{\gamma}$, is complex and spin dependent (originating from the spin-orbit coupling), where $\gamma \in \{x,y,z\}$ indicates the red, green and blue second-neighbor links as shown in Fig.\ref{lattice}(b). $\sigma^{\gamma}$ and $\sigma^{0}$ are the usual 2$\times$2 Pauli and identity matrices, respectively. The complex contribution of the second-neighbor hopping term (proportional to $t''$) is a result of hopping via $d$-$p$-$d$ ligands, while the real part is a result of direct overlap between $d$ orbitals, leading to the spin-independent amplitude $t'$.\cite{shitade:prl09}  Any nonzero value of $t''$ immediately opens a gap in the spectrum and turns the semi-metallic phase to a $\mathrm{Z_2}$ topological insulator phase.\cite{shitade:prl09} 

We include the Coulomb interactions by adding a Hubbard term,
\bea \label{Hubbard} 
{\cal H}=H_0+U\sum_{i}n_{i\uparrow}n_{i\downarrow}.
\eea 
The model \eqref{Hubbard} has been studied in the weak and intermediate interaction regime  by use of slave-spin theory.\cite{Ruegg:prl12} The quantum spin Hall insulator found in the weak interaction limit is unstable to a valence-bond solid (VBS), a very close relative of the expected AFM phase, for small spin-orbit coupling, and an exotic topological phase beyond a critical spin-orbit coupling strength, $t''$. Since the slave-spin theory is a variational approach, it cannot capture the possible magnetic phases which arise in the limit of  strong Coulomb interaction. 
In order to address this weakness of the slave-spin theory, we analytically take the limit of strong Hubbard interaction (at half-filling in the $j=1/2$ band) which results in the exchange Hamiltonian 
\bea 
\label{H}
 H=\sum_{i,j}J_{ij}\textbf{S}_i\cdot\textbf{S}_j-J_3\sum_{\ll ij\gg \gamma}(\textbf{S}_i\cdot\textbf{S}_j-2S^{\gamma}_iS^{\gamma}_j),\eea 
where $J_{ij}$=$J_1=4t^2/U$ and $J_{ij}$=$J_2=4(t')^2/U$ stands for first and second neighbor links, respectively, [both contained in the first term of \eqref{H}], and $J_3=4(t'')^2/U$ denotes the strength second neighbor contributions coming from the imaginary spin-dependent hopping term. The $S_i$ denotes the effective spin-1/2 moment of the $\mathrm{Ir}^{4+}$ ions, and the $\gamma$ indicates the direction of the links on the triangular lattice as described in Fig.\ref{lattice}(b). The model \eqref{H} is different from the known $J_{1}$-$J_2$-$J_3$ model with all isotropic exchange coupling previously studied in the literature\cite{Fouet:2001fk,Reuther2:prb11,Albuquerque:prb11} because in our model the third term explicitly breaks the $\mathrm{SU(2)}$ spin symmetry which will have important influences on the magnetic phases of the model, and because third-neighbor couplings are {\em not} considered. In the remainder of this paper we study the phase diagram of the Hamiltonian \eqref{H} by a combination of analytical methods and exact diagonalization on finite size clusters. 

\section{SCHWINGER BOSON MEAN FIELD THEORY 
\label{schwinger_sec}}
Recent numerical studies of the Hubbard model on the honeycomb lattice show a spin-gapped phase with no long-range correlations and no broken symmetries at an intermediate regime of Coulomb interaction $3.5$$<$$U/t$$<$$4.3$,\cite{Meng:nature10,Hohenadler:prl11,Hohenadler:prb12,Zheng:prb11,Wu:prb12} though more recent QMC results suggests the QSL phase may not be present at all in the Kane-Mele-Hubbard model.\cite{Sorella:1207.1783}
 In terms of a $J_1$-$J_2$ Heisenberg model on honeycomb lattice, this spin liquid phase is located around $J_2/J_1$$\approx$$0.06$.\cite{Yang:EPL11} The existence of a spin disordered phase was further confirmed by other techniques, including functional renormalization group\cite{Reuther2:prb11}, mean field and exact diagonalization,\cite{Albuquerque:prb11} but for higher values of $J_2/J_1$. One way wonder if the third term in Eq.\eqref{H}, which is anisotropic and has some degree of frustration, may help stabilize the spin liquid phase. To answer this question, we employ Schwinger boson mean-field theory to investigate the stability of the spin liquid phase. We find the coupling $J_3$ actually tends to stabilize a magnetically ordered phase instead of disordered one. This technique has proven successful in incorporating quantum fluctuations\cite{Ceccatto:prb93,Trumper:prl97} and has beed applied to the frustrated Heisenberg model on honeycomb,\cite{mattsson:prb94,cabra:prb11} kagome and triangular lattices.\cite{sachdev:prb92,wang:prb06}
 
In the Schwinger boson approach the spin operators are replaced by two flavors of bosons at each site,
\bea 
\label{boson_rep}
 \textbf{S}_i=\frac{1}{2}\textbf{b}_i^{\dag}\vec{\sigma}\textbf{b}_i,
 \eea 
 where $\textbf{b}^{\dag}=(b^{\dag}_{i\uparrow},b_{i\downarrow})$ are bosonic operators and $\vec{\sigma}$ is the vector of Pauli matrices. For this to be a faithful representation at each site, the following constraint should be imposed: \bea 
 \hat{n}_i=\sum_{\sigma}b^{\dag}_{i\sigma}b_{i\sigma}=2S.
 \eea 
 At the mean field level this constraint is imposed on average, namely $\langle\hat{n}_i\rangle=2S$, which is taken into account by Lagrange multiplier $\lambda$, taken to be independent of the site $i$. The exchange interactions can be written as follows, which make the model suitable for constructing a mean-field theory, 
\bea 
\label{exchange_decoupling}
 &&\textbf{S}_i\cdot\textbf{S}_j=\hat{\chi}^{\dag}_{0,ij}\hat{\chi}_{0,ij}-\hat{\Delta}^{\dag}_{0,ij}\hat{\Delta}_{0,ij},\nonumber \\
&&\textbf{S}_i\cdot\textbf{S}_j-2S^{x}_iS^{x}_j=\hat{\Delta}^{\dag}_{x,ij}\hat{\Delta}_{x,ij}-\hat{\chi}^{\dag}_{x,ij}\hat{\chi}_{x,ij},\nonumber \\ 
&&\textbf{S}_i\cdot\textbf{S}_j-2S^{y}_iS^{y}_j=\hat{\Delta}^{\dag}_{y,ij}\hat{\Delta}_{y,ij}-\hat{\chi}^{\dag}_{y,ij}\hat{\chi}_{y,ij},\nonumber \\ 
&&\textbf{S}_i\cdot\textbf{S}_j-2S^{z}_iS^{z}_j=\hat{\Delta}^{\dag}_{z,ij}\hat{\Delta}_{z,ij}-\hat{\chi}^{\dag}_{z,ij}\hat{\chi}_{z,ij}, 
\eea where the $\Delta$'s and $\chi$'s describe the paring and hopping of bosons. The full expressions are given in Appendix~\ref{bosonH}. 

We use mean field-theory as a variational approach and decouple the above expressions in different channels. Hence, the bosonic Hamiltonian becomes 
\bea 
H&=&J_1\sum_{<ij>}\chi^{\ast}_{0,ij}\hat{\chi}_{0,ij}-\Delta^{\ast}_{0,ij}\hat{\Delta}_{0,ij}+\mathrm{h.c.}\nonumber\\
&&+J_2\sum_{nnn}\chi^{\ast}_{0,ij}\hat{\chi}_{0,ij}-\Delta^{\ast}_{0,ij}\hat{\Delta}_{0,ij}+\mathrm{h.c.}\nonumber\\
&&-J_3\sum_{nnn,\alpha}\Delta^{\ast}_{\alpha,ij}\hat{\Delta}_{\alpha,ij}-\chi^{\ast}_{\alpha,ij}\hat{\chi}_{\alpha,ij}+\mathrm{h.c.}\nonumber\\
&&+\lambda\sum_{i}(\hat{n}_i-2S)+E_0,
\eea
where $E_0$ is an energy constant. 
As usual, the minimization of the ground state energy with respect to mean-field parameters provides a set of equations which should be solved self-consistently. Although it is possible to solve these equations, in practice it is a formidable task to find the solution. We therefore use the pairings and hoppings as variational parameters which can be tuned by the couplings $J_1,J_2,J_3$. Moreover, we should note that since the representation in Eq.\eqref{boson_rep} has a $\mathrm{U(1)}$ gauge redundancy, the mean-field ansatz should be invariant under a combined physical symmetry group and gauge-group operation which is called a projective symmetry group (PSG) operation.\cite{wen:QFTbook} In the PSG each physical symmetry is implemented followed by a particular gauge rotation such that the mean field ansatz is left invariant. We consider a uniform ansatz with zero-flux,\cite{wang:prb10} which would inspire a candidate for  the short-range resonance valence bond (RVB) state.\cite{sachdev:arxiv10} The ansatz is defined as 
\bea 
&&\chi_{0,ij}=\chi_{1},~~\Delta_{0,ij}=\Delta_{1},~~~ \mathrm{NN}\nonumber\\ 
&&\chi_{0,ij}=\chi_{2},~~\Delta_{0,ij}=\Delta_{2},~~~ \mathrm{NNN}\nonumber\\
 &&\chi_{\alpha,ij}=\chi_{\alpha},~~\Delta_{\alpha,ij}=\Delta_{\alpha},~~~ \mathrm{NNN}~~
~~~\alpha=x,y,z.
\eea 
Fourier transformed, the Hamiltonian can then be easily diagonalized as
\bea \label{H_k} H=\frac{1}{2}\sum_{k}\Psi^{\dag}_kH(k)\Psi_k+E_0,\eea where $\Psi_{k}^{T}=(b_{kA\uparrow}b_{kA\downarrow}b_{kB\uparrow}b_{kB\downarrow}b^{\dag}_{-kA\uparrow}b^{\dag}_{-kA\downarrow}b^{\dag}_{-kB\uparrow}b^{\dag}_{-kB\downarrow})$, and 
 \bea H(k)=\left(\begin{array}{ccc}h_k&\Delta_k\\ \Delta^{\dag}_{k}&h_{-k}\end{array}\right).\eea
For full expressions of $h_k$ and $\Delta_k$ see Appendix \ref{Hk}.
\begin{figure}[t]
\begin{center}
\includegraphics[width=8cm]{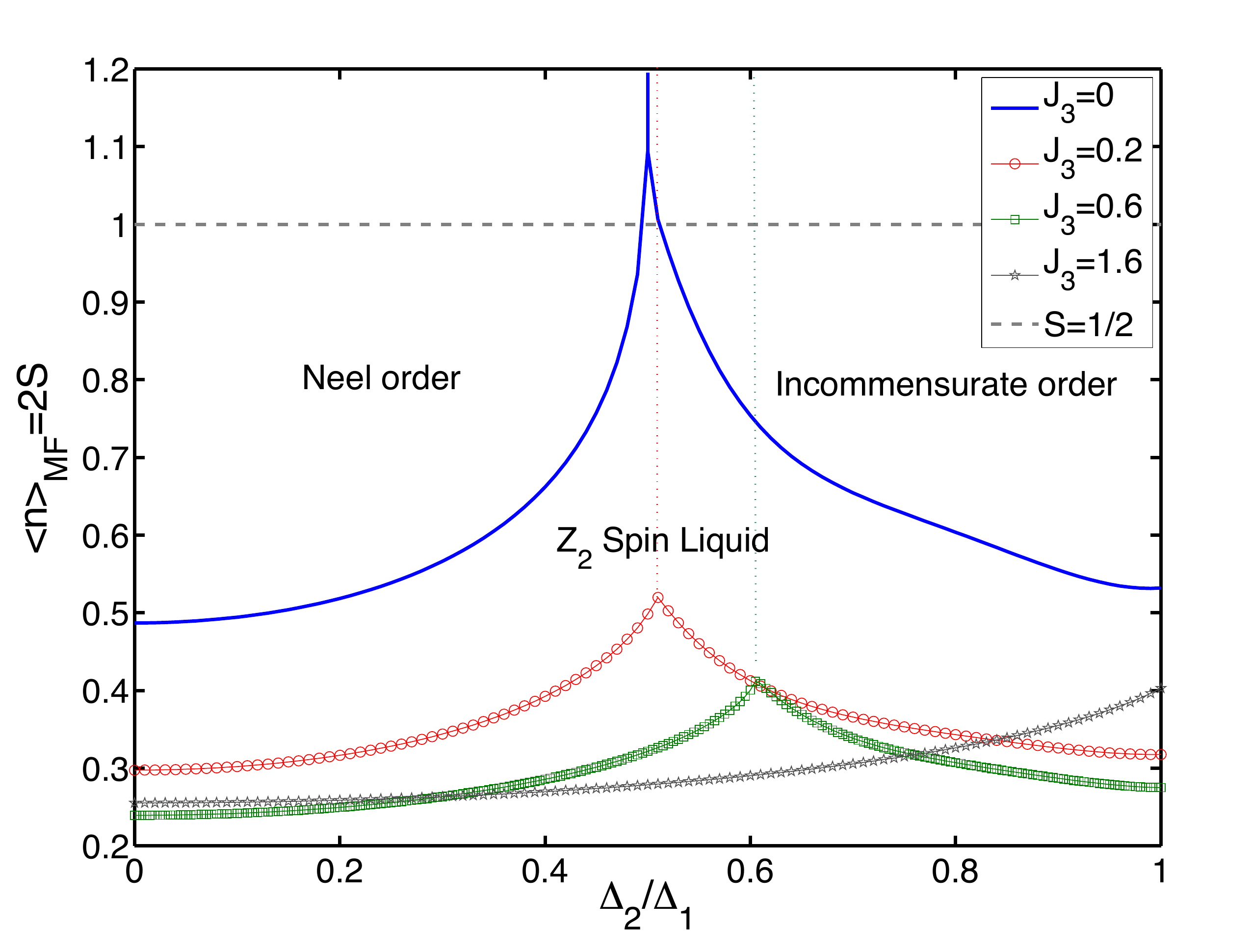}
\caption{(Color online) Schwinger Boson mean-field phase diagram of the model in Eq.\eqref{H}. The vertical axis is the average of boson density at each site, and the horizontal axis is the variational parameter $\Delta_2/\Delta_1$ which is a measure of the strength of second term in Hamiltonian. Different curves correspond to different values of $J_3$ ($\Delta_3=1$). The dashed line indicates the physical value of spin, namely $S=1/2$, which crosses the solid blue line in a very small window around $\Delta_2/\Delta_1=0.5$ indicating the existence of a $Z_2$ spin liquid phase. By increasing$J_3$, the corresponding red (circle), square (green) and grey (star) curves evade a crossing of the dashed line making the spin liquid phase very unlikely. Note that the region below each curve is a $\mathrm{Z_2}$ spin liquid phase, and above it is a magnetically ordered phase of either commensurate or incommensurate.} \label{schwinger_pd}
\end{center}
\end{figure}

With respect to the couplings $J_i$'s the bosons may condense at some wavevector in the Brillouin zone. The zeros in the energy spectrum of the condensate is used to determine the magnetic ordering that develops in the system.\cite{sachdev:prb92} On the other hand, a gapped spectrum of bosons could signal the existence of a disordered phase. Hence, we can determine the phase boundary between condensed and uncondensed regions, {\it i.e.} ordered and disordered phases. For simplicity we take $\Delta_{\alpha}=\Delta_3$. The phase diagram is shown in Fig.\ref{schwinger_pd}. Shown is a plot of the mean-value of the boson number at each site versus $\Delta_2/\Delta_1$ for different values of $\Delta_3/\Delta_1$. For $\Delta_3=0$ there is a narrow region around $\Delta_2/\Delta_1\approx 0.5$\cite{wang:prb10} which is crossed by the dashed line denoting the real value of $S=1/2$. In this region the bosonic spectrum is gapped, thus the system is spin disordered. Because of the second neighbor pairing which ``Higgs" the $\mathrm{U(1)}$ gauge bosons down to $\mathrm{Z_2}$, this spin liquid phase is stable\cite{wang:prb10} unlike the $U(1)$ spin liquid phase found in Ref.[\onlinecite{mattsson:prb94}]. Other parts of the phase diagram are magnetically ordered. At small values of $J_2$ the ordered phase is a Ne\'el phase where the bosons condense at the center of Brillouin zone, and for large values of $J_2$ the condensations occur at finite momenta which would lead to an incommensurate magnetically ordered phase.\cite{wang:prb10}    

Upon the inclusion of the third term $J_3$, the spin liquid window disappears, and as seen in Fig.\ref{schwinger_pd} the phase boundary is not crossed by the $S=1/2$ line anymore. Therefore, it seems that the anisotropic $J_3$-term strongly suppresses quantum fluctuations and favors a magnetically ordered state. This is in contrast to the $J_1$-$J_2$-$J_3$ model, where the third-neighbor term was shown to stabilize a spin liquid phase via the same Schwinger boson method.\cite{cabra:prb11}  (Recall that our $J_3$ is a second-neighbor coupling coming from $t''$ in Eq.\eqref{H0}.) Despite the apparent breaking of $\mathrm{SU(2)}$ symmetry by the $J_3$-term in Eq.\eqref{H}, it has a ``hidden" rotational invariance which could stabilize a magnetically ordered phase. We will elaborate on this point in the following section.
\begin{figure}[t]
\begin{center}
\includegraphics[width=8cm]{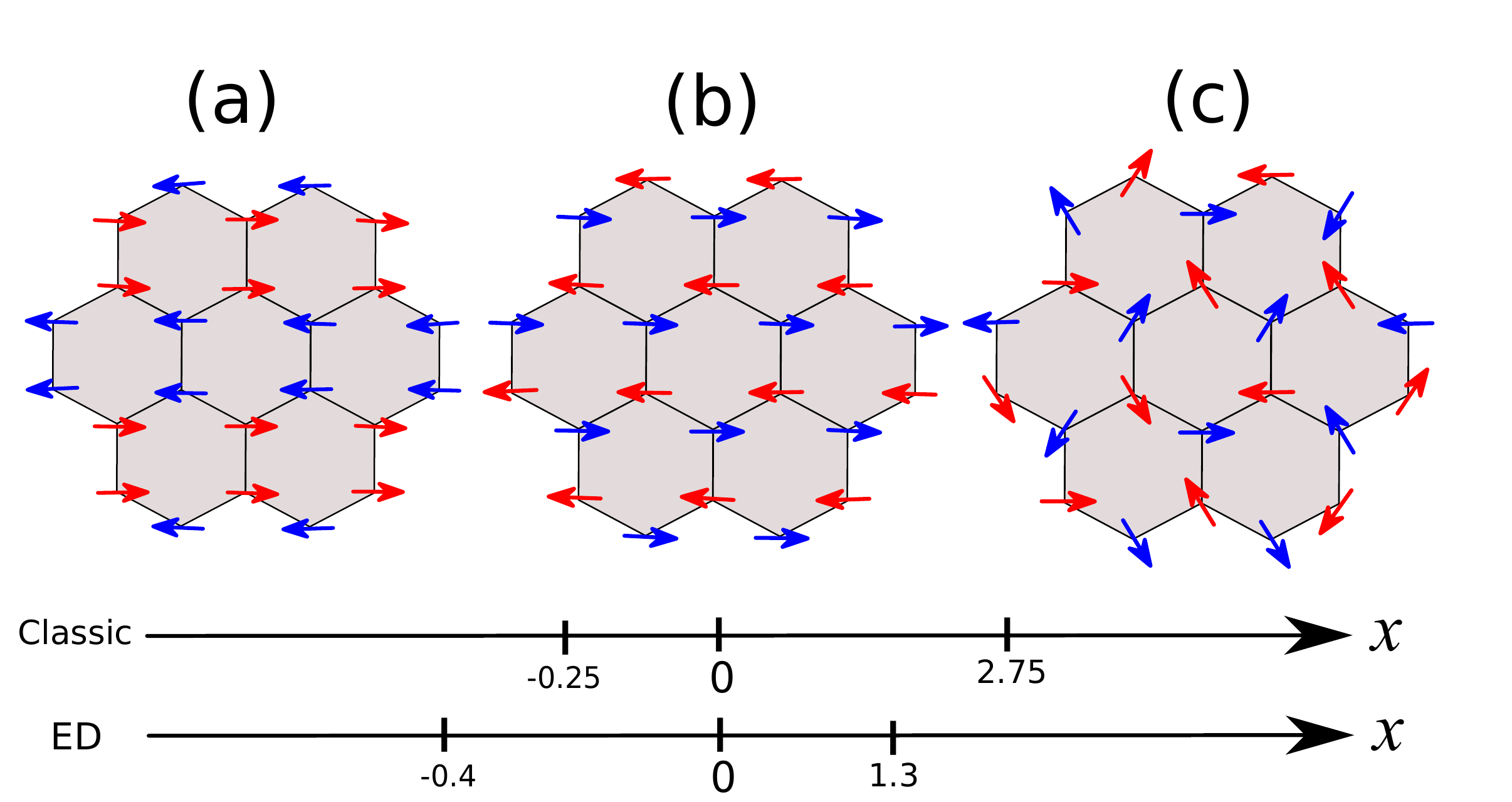}
\caption{(Color online) Top: Classical spin ordering of the $J_1$-$J_3$ model \eqref{H} with $J_2$=0. (a) Stripy order for $J_3 < x_{c1}$, (b) Ne\'el order for $x_{c1} <J_3<x_{c2}$, and (c) Spiral order for $J_3 > x_{c2}$. Bottom: The phase diagram as a function of $x=J_3/J_1$ and the positions of phase transition points $x_{c1}$ and $x_{c2}$ for the classical and quantum model. The results for the quantum model were obtained by exact diagonalization (ED) on finite size clusters.} \label{ordering}
\end{center}
\end{figure}
\section{Candidate magnetically ordered phases\label{condicate_phases}}
The discussion in the preceding section showed that the $J_3$ term tends to stabilize ordered phases, and therefore disfavors spin liquids. In this section we discuss possible magnetically ordered phases at the limit of large spin values, namely the classical orders. Ground-state configurations are readily obtained in some limiting cases. Since the model with $J_3=0$ has been studied before,\cite{Katsura:JSP86} here we focus on the $J_1$-$J_3$ model, and set $J_2=0$. We will discuss the effect of the $J_2$ coupling in Sec.\ref{conclusion}.  In the limit of vanishing $J_3$ the magnetic phase is the usual Ne\'el phase. On the other hand, in the limit of $|J_3|$$\gg$$J_1$, the model is decoupled to two trianglular latticies each governed by the $J_3$ term, namely $ H_{tri}=-J_3\sum_{<ij>\gamma}\textbf{S}_i\cdot\textbf{S}_j-2S^{\gamma}_iS^{\gamma}_j,$ where, now $<ij>$ stands for NN links on the triangular lattice. This Hamiltonian, despite being obtained in an extreme limit, gives a proper low energy description of the cobaltates where the spin-orbit coupling is much stronger than the superexchange coupling.\cite{Khaliullin:PTPS05} The Hamiltonian $H_{tri}$ is not frustrated as can be seen by dividing the triangular lattice to four sublattices and performing the following local transformations on the four sublatticies shown in Fig.\ref{lattice}(b) by empty and blue circles, squares and triangles:
\bea \label{transformation}&&\mathrm{Empty~circle}:~~\tilde{S}^{x}=S^x,~ \tilde{S}^{y}=S^y,~\tilde{S}^{z}=S^z,\nonumber \\ 
&&\mathrm{Red~triangle}:~~\tilde{S}^{x}=S^x,~ \tilde{S}^{y}=-S^y,~\tilde{S}^{z}=-S^z,\nonumber \\
&&\mathrm{Green~square}:~~\tilde{S}^{x}=-S^x,~ \tilde{S}^{y}=S^y,~\tilde{S}^{z}=-S^z,\nonumber \\
&&\mathrm{Blue~circle}:~~\tilde{S}^{x}=-S^x,~ \tilde{S}^{y}=-S^y,~\tilde{S}^{z}=S^z.\nonumber \\\eea

In the transformed basis the Hamiltonian becomes fully isotropic,
 \bea 
 \label{Htri} 
 H_{tri}=J_3\sum_{\ll ij\gg}\tilde{\textbf{S}}_i\cdot\tilde{\textbf{S}}_j.
 \eea
Thus, the ground state will be the well known $\mathrm{120^{o}}$ ordering on the triangular lattice for $J_3$$>$$0$ and a fully ferromagnetic state for $J_3$$<$$0$. Transforming back to the original spins, we obtain stripy (or zig-zag) order for latter case and spiral order for the former case as shown, respectively, in Fig.\ref{ordering}(a) and Fig.\ref{ordering}(c). The stripy order was also observed in the Kitaev-Heisenberg model.\cite{Chaloupka:prl10} Note that the stripy order and spiral order result from the explicit $\mathrm{SU(2)}$ spin symmetry breaking. Therefore, any phase transition from the Ne\'el phase to these phases  should be related to a fully broken spin rotational symmetry, as such a phase transition is absent in the Kane-Mele-Hubbard model which preserves $\mathrm{U(1)}$ spin symmetry.\cite{Hohenadler:prl11,Reuther:prb12} 

Having established the orderings in the extreme limits of the model, we can now estimate the classical transition points by comparing the ground state energy per spin:    
\bea &&E_{Neel}=\frac{E_0}{2J_1S^2N_{cell}}=-\frac{3}{2}-x,\nonumber \\
&&E_{stripy}=\frac{E_0}{4J_1S^2N_{cell}}=-\frac{1}{2}+3x,\nonumber \\
&&E_{spiral}=\frac{E_0}{24J_1S^2N_{cell}}=-\frac{1}{8}-\frac{3}{2}x,\nonumber \\ \eea 
where $x=J_3/J_1$.  For $x<-0.25$ the stripy phase is favored, for $-0.25<x<2.75$ the Ne\'el phase, and for $x>2.75$ the commensurate spiral phase is favored. In the next section we will show that the quantum effects shift the phase boundaries.     

\section{exact diagonalization on finite size lattices\label{lanczos_sec}}
In this section we use Lanczos exact diagolalization (ED) on finite size lattices to locate the critical points and identify different magnetic phases. We consider two-dimensional lattices with N=13,17,19,24 sites. For N=24 we considered both periodic and open boundary conditions. Note that the periodic lattice with N=24 is the minimum lattice size whose extra frustration due to the boundary of the lattice is avoided. This cluster is shown in Fig.\ref{lattice}(b). The N=24 lattice avoids extra frustration from the periodicity of transformations in Eq.\eqref{transformation}, which is 2, because it is consistent with the periodicity of magnetic ordering of Eq.\eqref{Htri} for $J_3>0$, which is 3 for each direction on the triangular lattice. Naively, if we take a lattice based on the primitive unite vectors of the honeycomb lattice, it will be of size 72, which is not practical for ED. For sizes smaller than 24 the open boundary condition should be worked out.
\begin{figure}[t]
\begin{center}
\includegraphics[width=8.5cm]{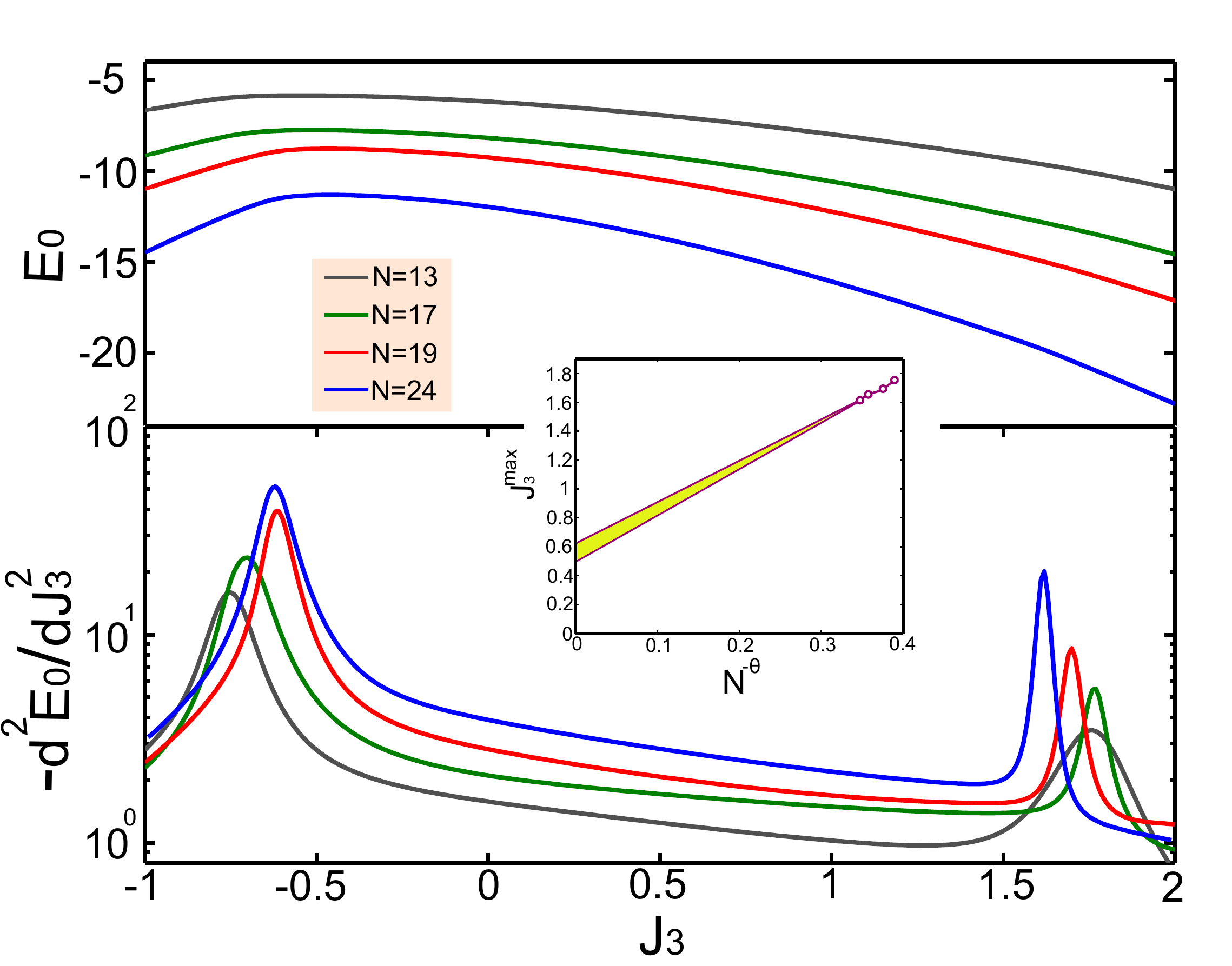}
\caption{(Color online) Upper panel: The ground state energy per site versus coupling $J_3$ for different lattice sizes, from top to bottom: N=13 (grey), 17 (green), 19 (red) and 24 (blue). All clusters used open boundary conditions, and we set $J_1$=1 and $J_2$=0. Lower panel: The second derivative of the ground state energy versus coupling. Note the vertical axis is in a logarithmic scale. Two clear peaks become more pronounced as the system size increases. For the largest lattice size the left peak appears around $J_3$=-0.62$\pm$0.01 and right one around $J_3$=1.62$\pm$0.01. Inset: finite size scaling to locate the approximate thermodynamic critical point with $\theta \approx 0.33$.} \label{gse_derivative}
\end{center}
\end{figure}

To find magnetic phases, a set of  ordered magnetic phases found classically in Fig.\ref{ordering} serves as a useful guide. The idea is to define the corresponding order parameter for the stripy (zig-zag) and Ne\'el phases. The unfrustrated model at $J_3$=0 possesses antiferromagnetic long-range order characterized by a staggered magnetization, with the staggered moment $m$=0.2677, which has been significantly reduced from classical moment by quantum fluctuations.\cite{Castro:prb06} This has been verified by various methods including spin-wave theory, the coupled cluster method, exact diagonalization, series expansions around the Ising limit, tensor network studies, variational Monte Carlo and quantum Monte Carlo (QMC) simulations (see A.F.Albuquerque, $\mathit{et~al}$\cite{Albuquerque:prb11} and references therien). For numerical purposes, a good order parameter of the Ne\'el phase would be the staggered magnetization squared on entire lattice,\cite{Albuquerque:prb11} 
\bea 
\label{m_neel} 
m^2_{Neel}=\frac{1}{S(S+1)}\left (\sum_{i}(-1)^{i}\textbf{S}_{i}\right)^2,
\eea 
where $S=N/2$. In the following, this staggered magnetization is used to determine the stability of the Ne\'el phase. At $J_1$=0 the model is also unfrustrated thanks to the transformation in Eq.\eqref{transformation}. As pointed out in preceding section for $J_3$$<$0 ($>$0), the model reduces to ferromagnetic (antiferromagntic) exchange coupling on two isolated triangular lattices, which can then described by the Hamiltonian in Eq.\eqref{Htri}. Therefore, for $J_3$$<$0 one can simply compile the following total ferromagnetic moment for rotated spins,
\bea 
\label{m_ferro} 
m^2_{stripy}=\frac{1}{S(S+1)}\left (\sum_{i}\tilde{\textbf{S}}_{i}\right)^2.
 \eea 
These order parameters and their vanishing (to within finite size limitations) describe the stability of the collinear ordered phases and their transition to a spiral phase.   

In Fig.\ref{gse_derivative}, we plot ground sate energy (upper panel) and its second derivative versus coupling $J_3$ for different lattice sizes. The energy shows a smooth behavior with maximum around $J_3$=-0.62$\pm$0.01 for the largest size N=24 with open boundary conditions. This maximum indicates that there is a phase transition at this coupling. The existence of a rather sharp peak appeared in the second derivative of the ground state energy (lower panel) clearly signals this phase transition. While it is rather hard to locate another phase transition just by looking at the behavior of energy, its second derivative shows the approximate location of the second-order transition to yet another phase. It occurs around $J_3$=1.62$\pm$0.01 for N=24. One can also see that by increasing the size of the lattices the peaks in the second derivative of the energy gets more pronounced. These critical values could be compared with the classical values in Fig.\ref{ordering} which show that quantum fluctuations significantly shift them to lower values. 

Yet, more precise location of critical points could be probed by finite size scaling. The maximum location of second derivative of energy scales as $J_3^{max}=J_3^{c}+aN^{-\theta}$ with system size. Our best fitting to available data as shown in inset of Fig.\ref{gse_derivative} indicate the thermodynamic phase transition occurs at significantly lower values somewhere in 0.47$<$$J_3^c$$<$0.65, and the exponent is $\theta \approx 0.33$. This approximate estimation is consistent with the critical coupling found at $\tilde{\lambda}_c=0.53$ in Ref.[\onlinecite{Reuther:prb12}]. We also used the similar ansatz of finite size scaling to locate the thermodynamic phase transition found for $J_3$$<$0, which shows that the critical point should be located in -0.3$<$$J_3^c$$<$-0.1 with exponent $\theta \approx 0.4$.   
\begin{figure}[t]
\begin{center}
\includegraphics[width=9cm]{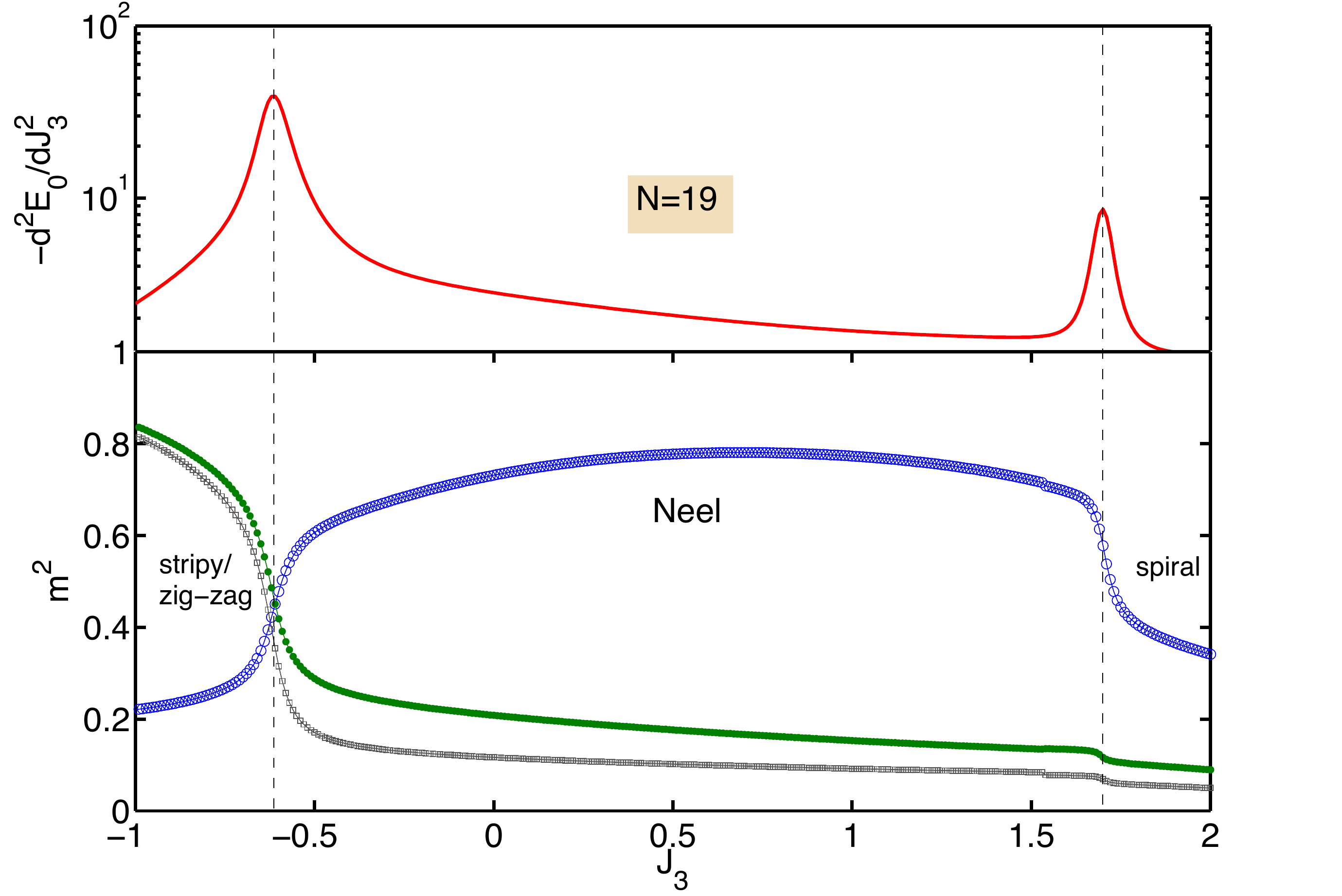}
\caption{(Color online) Upper panel: Second derivative of the ground state energy per site versus coupling $J_3$ for a lattice with N=19 and open boundary conditions. We set $J_1$=1 and $J_2$=0. Lower panel: Order parameters squared $m^2_{Neel}$ (empty blue circles), $m^{2}_{stripy}$ (green solid dots) and $m^{2}_{zig-zag}$ (squares). At the left critical point $J_3$=-0.62 the stripy oder reduces and the Ne\'el order starts to increase. Both orders decrease across the right critical point around $J_3$=1.62 where a spiral order begins to develop. The vertical dashed lines are guides to the eyes.} \label{N19}
\end{center}
\end{figure}

\begin{figure}[t]
\begin{center}
\includegraphics[width=9cm]{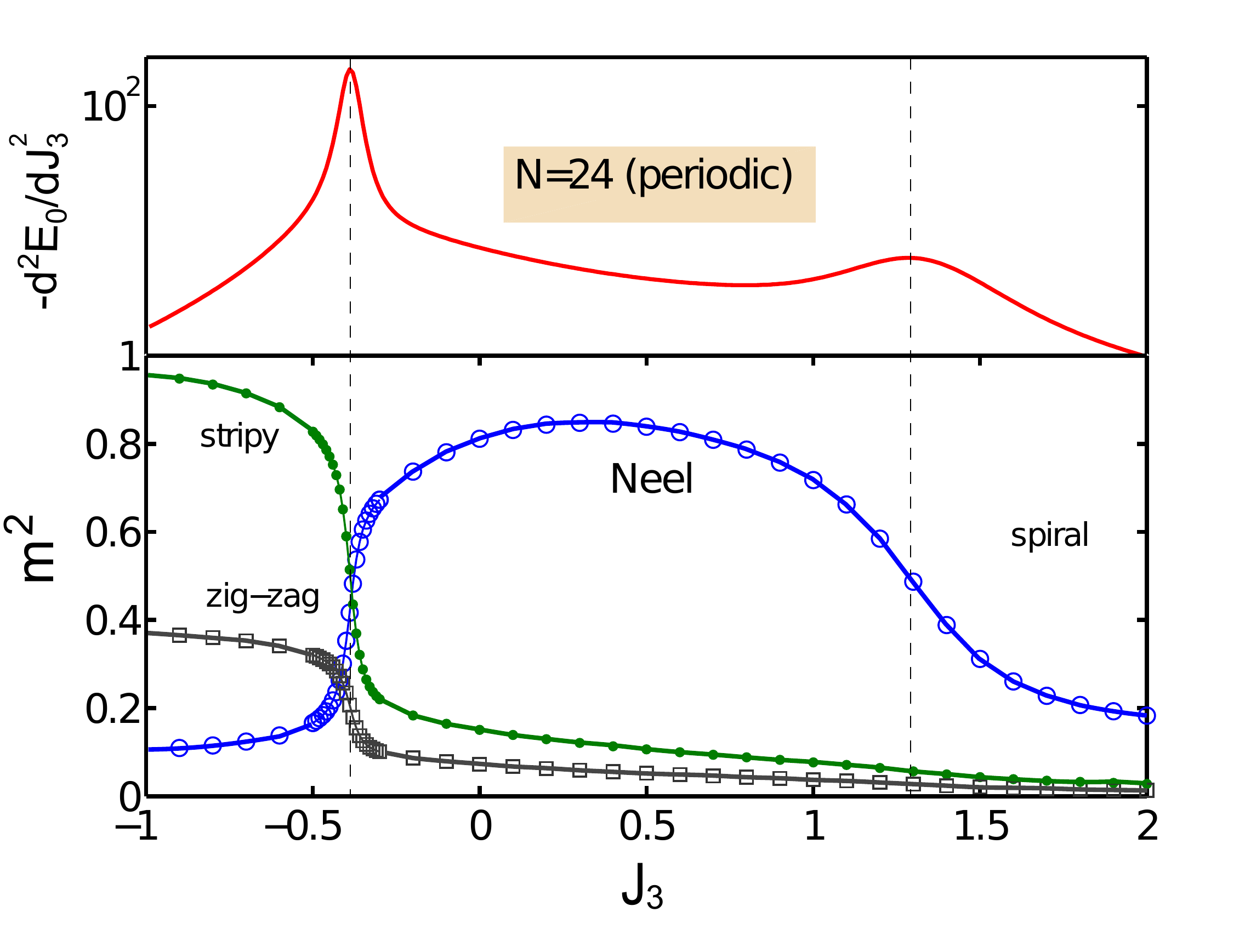}
\caption{(Color online) The same quantities are plotted as in Fig.\ref{N19} but for N=24 with periodic boundary conditions. Note that the approximate positions of critical points have been changed.} \label{N24}
\end{center}
\end{figure}

The nature of magnetic phases around the critical points can be determined by examining the order parameters in Eq.\eqref{m_neel} and Eq.\eqref{m_ferro}. Their behaviors are elaborated for N=19 and N=24 in Fig.\ref{N19} and Fig.\ref{N24}, respectively. For N=24 we presented the results for periodic boundary condition to avoid boundary effects. Besides the order parameters we also included the second derivative of the energy to clearly show the singularity in the derivative of the energy coincides with the drop or onset of various order parameters. The nature of the ordered phases is already clear for the small size cluster N=19. For values of $J_3$$<$-0.62 the dominant order is the stripy order (See Fig.\ref{ordering}(a)) which is steeply reduced across the critical point, where the Ne\'el phase starts to develop over the intermediate range of coupling. At the fully isotropic point $J_3$=0 the manetization squared is $m^2$=$0.81/4$$\approx$0.2 in agreement with result in Ref.[\onlinecite{Albuquerque:prb11}]. The Ne\'el order parameter increases even beyond the isotropic $J_3$=0 point owing to the ferromagnetic nature of the isotropic term in Eq.\ref{H}. These two collinear phases then drop down across the second critical point, with the onset of spiral order. 

As far as the magnetic structure of the layered Iridate $\mathrm{Na_2IrO_3}$ is concerned, recent resonant x-ray scattering\cite{liu:prbR11} showed magnetic Brag peaks at wave vectors consistent with both stripy and zig-zag orders. However, first principles calculations\cite{liu:prbR11} and very recent inelastic neutron scattering (INS) experiments\cite{choi:prl12,Ye:prbR12,singh:prl12} showed that further-neighbor exchange interactions are strong, which in turn makes the zig-zag configuration the most likely magnetic order. 

Similar to the Heisenberg-Kitaev model, the model in Eq.\eqref{H} can not explain the zig-zag order as its ground state.  However, we would like to argue that even in the context of the model in Eq.\eqref{H}, we have some degree of zig-zag ordering. The argument traces back to the Hamiltonian of transformed spins on triangular lattice in Eq.\eqref{Htri} for $J_3$$<$0 leading to ferromagnetic order of rotated spins. Transforming back to the original spins, the ordering on honeycomb lattice will be stripy or zig-zag depending on the sign of the NN coupling: stripy for $J_1$$>$0 and zig-zag for $J_1$$<$0. Often $J_1$$>0$, so the former phase is energetically favored. But it seems quantum fluctuations make them close to each other in the strength of their respective order parameters. To elaborate this, in bottom panel of Fig.\ref{N19} we also plot the zig-zag order parameter of the ground state. It is clearly seen that, despite having higher energy than the stripy phase, the ground state posses a high degree of zigzag ordering. Note that this holds at the level of an exchange model with up to second-neighbor coupling unlike the well known $J_1$-$J_2$-$J_3$\cite{Fouet:2001fk,cabra:prb11} or Kitaev-Heisenberg-$J_2$-$J_3$\cite{Kimchi:prbR11} model. The results for N=24 with periodic boundary conditions are shown in Fig.\ref{N24}. The overall features of the plot is the same as  Fig.\ref{N19} except the approximate positions of the critical points have been displaced: The transition from the stripy to Ne\'el and from Ne\'el to the spiral phase would occur around $J_3$=-0.4 and $J_3$=1.3, respectively. Moreover, the value of zig-zag order is still high with a ratio $m_{zig-zag}/m_{stripy}$=0.62 at $J_3$=-1.      

\section{fidelity analysis of the phase transitions\label{fidelity_sec}}            
In this section we use a quantum information theoretic tool called fidelity to analyze the quantum phase transitions described in the preceding sections. For pure states, say $|\psi_{1}\rangle$ and $|\psi_{2}\rangle$, it simply measures the overlap between them as $F=|\langle \psi_1|\psi_2\rangle|$, and so is a measure of distinguishability between the states. The states $|\psi_{1}\rangle$ and $|\psi_{2}\rangle$ could be ground states of the Hamiltonian in Eq.\eqref{H} corresponding to slightly different values of tuning parameter, say $J_3$ and $J_3$+$\delta$, namely $F(J_3,\delta)$=$|\langle\psi(J_3)|\psi(J_3+\delta)\rangle|$, where $\delta$ is a small quantity. Naturally for the same (orthogonal) states it will be unity (zero). While far away from the critical points the distinguishability is not significant, across a phase transition there is a drastic change in the fidelity as the ground states on different sides of critical points are totally different. Hence, the fidelity could be a strong signature of a quantum phase transition.\cite{Zanardi:pre06,gu:InjpB08} Moreover, its scaling is intrinsically connected to derivatives of the ground state energy, and it was argued that the fidelity susceptibility (FS) defined as $S(J_3)=\partial^{2}_{\delta}F(J_3,\delta)|_{\delta=0}$=$2\lim_{\delta\rightarrow 0}\frac{1-F(J_3,\delta)}{\delta^2}$ is a more sensitive tool than the second derivative of the ground state energy to detect the critical points.\cite{chen:pra08}  In particular, close to criticality,  
\bea 
-\frac{\partial^2E_0}{\partial J^{2}_3}\sim \frac{1}{\Delta}, \nonumber \\ 
S(J_3)\sim\frac{1}{\Delta^2},
  \eea 
  where $\Delta$ is the energy gap. This relation explicitly indicates that a singularity in the second derivative of the energy will imply a singularity in the FS, and the singularity in the FS is more pronounced close to phase transition. Moreover, as the definition of the fidelity or the FS does not rely on the notion of either a local order parameter or symmetry, it has been used to detect topological phase transitions which evade a description based on a local order parameter.\cite{Abasto:pra08,Langari:njp12}
\begin{figure}[t]
\begin{center}
\includegraphics[width=9cm]{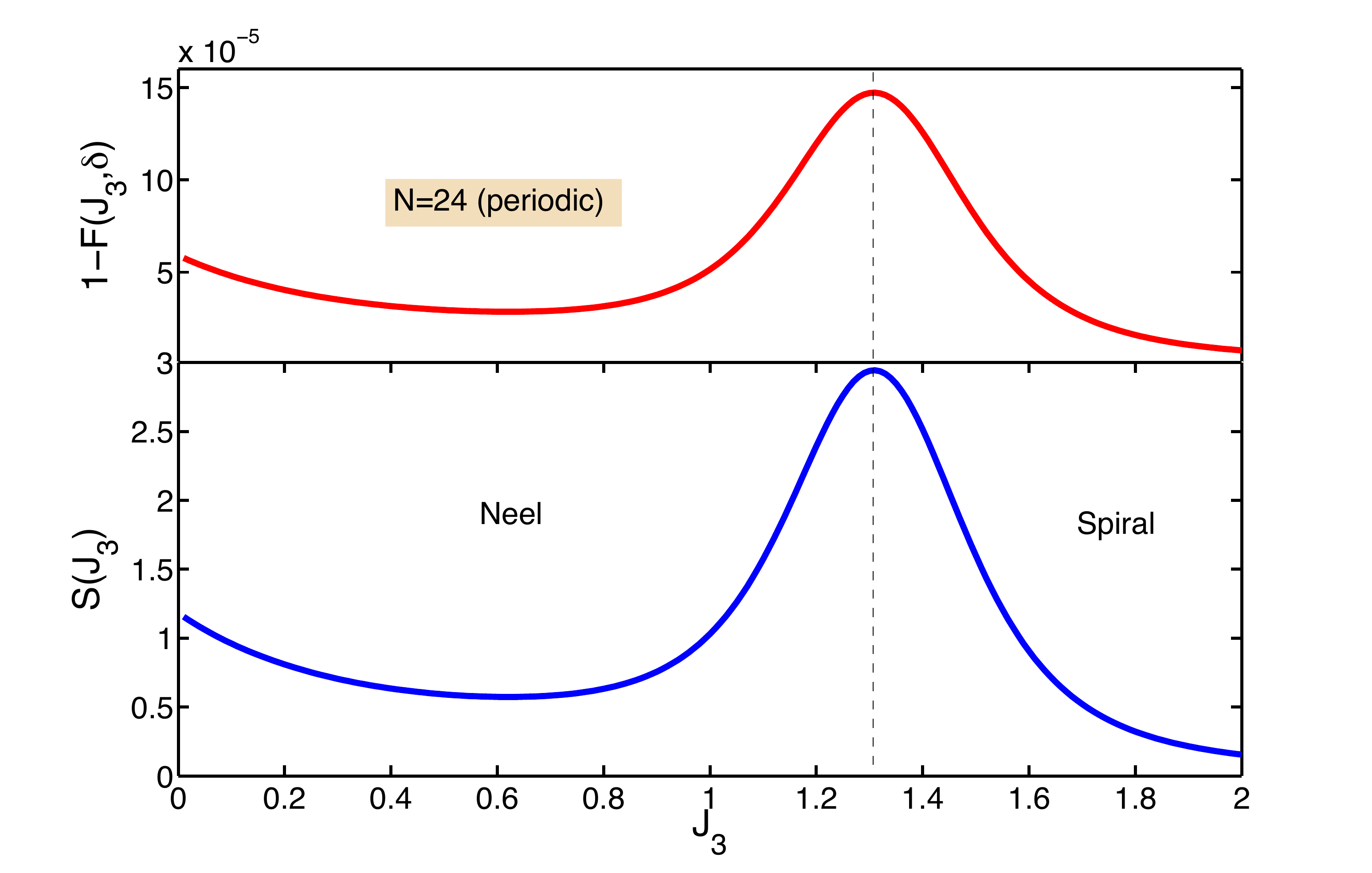}
\caption{(Color online) Fidelity (upper panel) and fidelity susceptibility (bottom panel) versus $J_3$ (with $J_2$=0) for N=24 with periodic boundary conditions and $\delta$=0.01. The singularity in S($J_3$) is much more pronounced than the singularity in the second derivative of ground state energy shown in Fig.\ref{N24}. The dashed line indicates the location of critical point $J_3$=1.3.}\label{fidelity}
\end{center}
\end{figure}

In Fig.\ref{fidelity}, we plot both fidelity (upper panel) and fidelity susceptibility (bottom panel) for positive values of $J_3$ and for a lattice size N=24 with periodic boundary conditions. Note that in upper panel we plotted 1-$F(J_3,\delta)$. Indeed, for $J_3$$<$0 the phase transition is already clear from the derivative of energy as shown in Fig.\ref{N24}. Of course, it is also clearly signaled in fidelity (not shown here). But the singularity in derivative of energy around $J_3$=1.3 is rather weak and has a bump-like behavior which make the precise determination of the location of the critical point difficult. Fidelity, as expected, exhibits a significant change in the ground state wavefunction across the critical point. Additionally, fidelity susceptibility reveals a more pronounced singularity at the critical point $J_3$=1.3 in contrast to the second derivative of energy in Fig.\ref{N24}.

\section{conclusions and further aspects\label{conclusion}}
In this work we studied the strong interaction limit of the Hubbard model in Eq.\eqref{Hubbard} whose noninteracting ground state is a two-dimensional topological band insulator fully breaking the $\mathrm{SU(2)}$ spin symmetry,\cite{shitade:prl09} which in turn makes the exchange couplings highly anisotropic and frustrated. Since the Schwinger boson mean-field theory shows that the anisotropic term favors magnetic ordered phases rather than a spin liquid phase which could arise in frustrated models, we first considered the $J_1$-$J_3$ model and set $J_2$=0. The results of exact diagonalization studies on finite size lattices are summarized in Figs.\ref{gse_derivative}, \ref{N19}, \ref{N24}, where it was shown that varying coupling $J_3$ (with $J_2$=0) stabilizes the stripy, Ne\'el, and spiral phases. For the largest lattice we considered, N=24, the critical points are at $J_3$=-0.4 and $J_3$=1.3 (with extrapolation to thermodynamic limit giving rise to $J^{c}_3$=-0.2$\pm$0.1 and $J^{c}_3$=0.55$\pm$0.1, respectively, by finite size scaling) corresponding to stripy-Ne\'el and Ne\'el-spiral transitions, respectively. The latter phase transition has also been reported in recent work.\cite{Reuther:prb12}  We further exploited the fidelity susceptibility to locate the latter phase transition more precisely. The appearance of stripy and spiral orders, and the phase transitions out of a Ne\'el phase are ascribed to the explicitly broken spin symmetry.    

These magnetic phase transitions can be used to shed light on correlation effects in two-dimensional topological band insulators. Indeed, the topological band insulator persists up to intermediate strengths of an on-site Hubbard interaction. However, the physics at intermediate and strong interactions is significantly influenced by strength of the spin-orbit coupling, $t''$, which determines the gap of the non-interacting topological insulator. For intermediate values of the Hubbard interaction, small and large values of $t''$ result in AFM (VBS) and QSH* phases, respectively.\cite{Ruegg:prl12} While the former breaks the time reversal symmetry, the latter still preserves time reversal symmetry with protected {\em collective} edge modes and bulk topological degeneracy. Our work based on the exchange Hamiltonian in Eq.\eqref{H} further revealed that this topological phase eventually breaks down at strong interacting limit to a magnetically ordered phase with a spiral texture. However, it appears that a phase transition as a function of spin-orbit coupling still persists in the large interaction limit.          
\begin{figure}[t]
\begin{center}
\includegraphics[width=8cm]{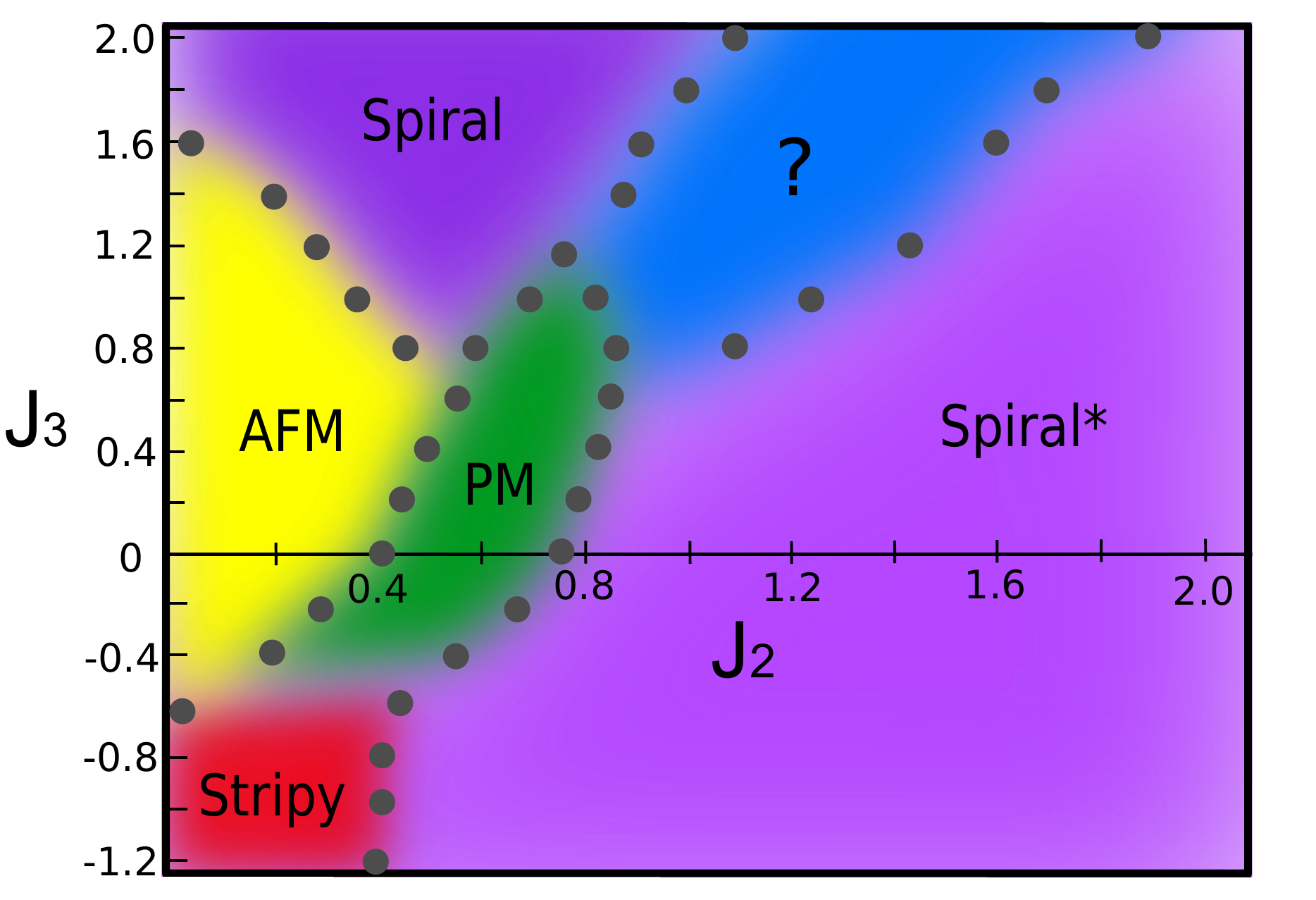}
\includegraphics[width=8cm]{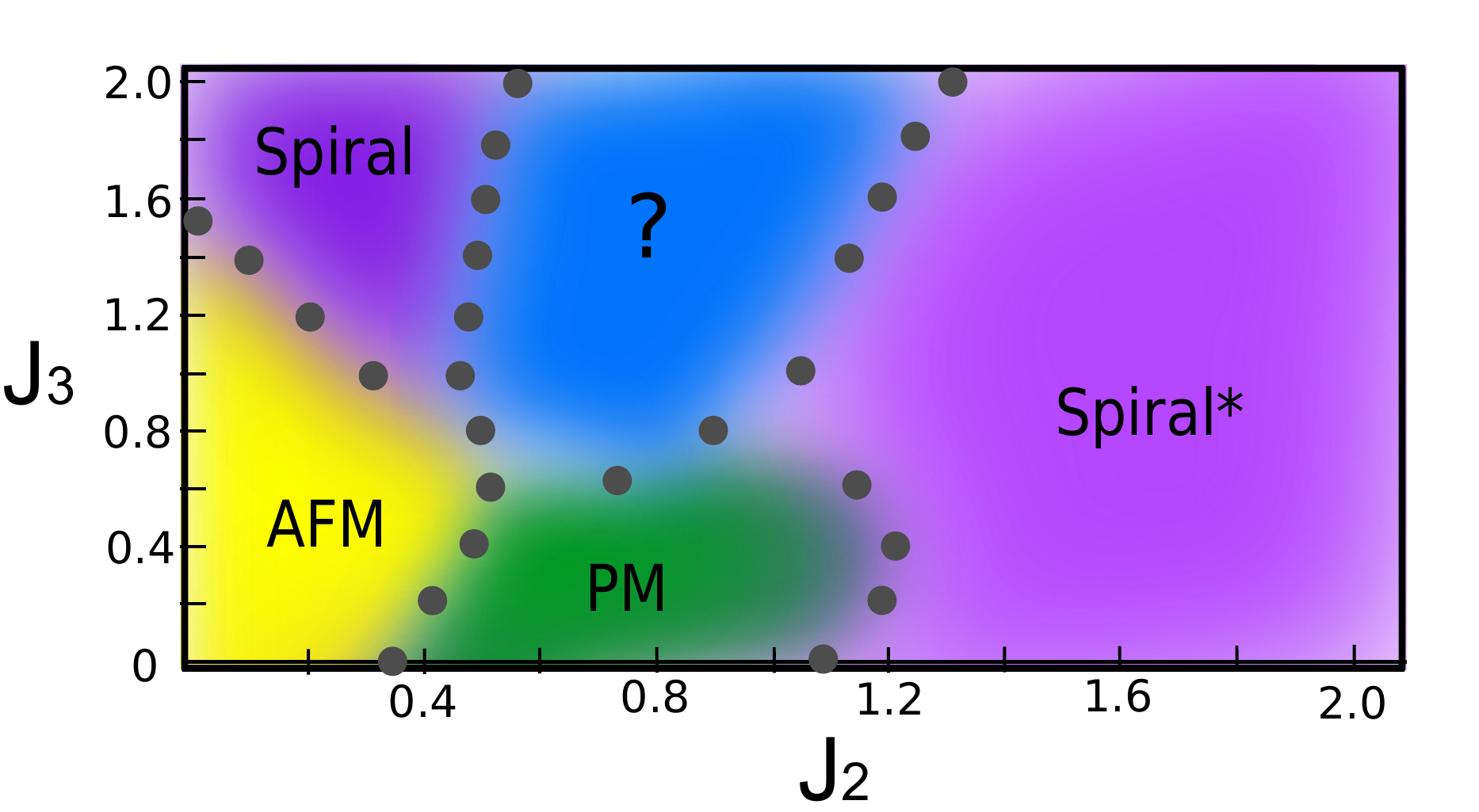}
\caption{(Color online) Schematic phase diagram of the model in Eq.\eqref{H} for a finite lattice with N=19 (top panel) and N=24 (bottom panel for $J_3$$>$0). Solid circles correspond to approximate positions of singularities in the derivative of the ground state energy. Phases are antiferromagntic (AFM), paramagnetic (PM), Stripy, Spiral and Spiral*. The nature of the phase marked by question sign needs further study.} \label{PDJ2J3}
\end{center}
\end{figure}

Thus far, we have considered the limit of $J_2$=0. In the rest of this section we discuss the affect of this isotropic exchange coupling term and map out the full phase diagram for a lattice sizes N=19 and N=24. The phase diagram in the $J_2$-$J_3$ plane is shown in Fig.\ref{PDJ2J3}.  To obtain it, we used knowledge of the phases appearing in various limits to distinguish different phases and the transitions between which are signaled by a singularity in the derivative of ground state energy. Although these lattice sizes are too small to locate the precise position of phase boundaries, we showed in preceding sections that enlarging the lattice just shifts the phase boundary somewhat. So, we believe that even systems of these small sizes can give a qualitative sense of the phase diagram of the model. 

On the $J_3$=0 axis the existence of  AFM, paramagnetic (PM) and spiral orders have been investigated in the literature.\cite{Reuther2:prb11,Albuquerque:prb11} The only difference with the classical phase diagram\cite{Katsura:JSP86,Fouet:2001fk} is the appearance of a PM phase at intermediate values of $J_2$. There are two spiral phases in phase diagram. The spiral phase on the $J_3$=0 axis should be distinguished from the spiral phase arising at the large values of $J_3$ ($J_2$=0), which is why we showed it with a star. At the extreme limits where either $J_2$$\gg$$J_1$ or $J_3$$\gg$$J_1$ the spiral* and spiral phases are adiabatically connected to the $\mathrm{120^{o}}$ ordering of, respectively, original and rotated spins (see Eq.\eqref{transformation}) on decoupled triangular lattices. Note that these two phases are separated by phase transitions and some phases in between and cannot be connected adiabatically on the $J_2$-$J_3$ plane.

The phase diagram in Fig.\ref{PDJ2J3} should be compared with the phase diagram reported for the isotropic $J_1$-$J_2$-$J_3$ model.\cite{Reuther2:prb11,Albuquerque:prb11} Note that in the latter model $J_3$ coupling stands for third-neighbor couplings on the honeycomb lattice. While this isotropic term stabilizes the AFM phase, the anisotropic coupling in Eq.\eqref{H} stabilizes the spiral order instead. However, in both models the PM phase persists up to intermediate coupling $J_3$. There is still one phase labeled by a question mark in Fig.\ref{PDJ2J3} which needs further study, being beyond the scope of this work. However, we can still argue which model may describe this undetermined phase. It appears this phase is stabilized when $J_2$$\sim$$J_3 \gtrsim J_1$. In this limit the second neighbor isotropic term in Eq.\eqref{H} becomes small and we will end up with a Kitaev-Heisenberg-type model on two triangular lattices coupled antiferromagnetically through $J_1$: \bea \label{Hk} H_k=J_1\sum_{<i,j>}\textbf{S}_i\cdot\textbf{S}_j+J_H\sum_{\ll ij\gg}\textbf{S}_i\cdot\textbf{S}_j+J_K\sum_{\ll ij\gg \gamma}S^{\gamma}_iS^{\gamma}_j, \nonumber \\ \eea where $J_H$=$J_2$-$J_3$ and $J_K$=$2J_3$. Last term describes Kitaev-type model on triangular lattice. Unlike the original Kitaev model\cite{Kitaev20032} which is exactly solveable due to trivalent structure of the honeycomb lattice giving rise to a bilinear representation of the Hamiltonian in term of Majorana fermions, an extension of the model to a triangular lattice is no longer trivial for spin-1/2 degrees of freedom. Even if we assume the ``parallel" Kitaev models with isotropic couplings on the triangular lattice are spin disordered, the coupling between layers, though small, may stabilize a magnetically ordered phase. It is unclear what this order may be. Recently it was shown that in the absence of coupling between lattices, namely $J_1$=$0$, the isotropic coupling of spins within each triangular lattice with $J_H$ gives rise to some non-coplanar spin orderings with nontrivial vortex structure.\cite{daghofer:1209.5895} But how $J_1$ coupling between triangular lattices might affect such an ordered state needs further investigation. We leave this question for future study.     

\section{acknowledgements}
We would like to thank Simon Trebst, Victor Chua, Fa Wan, Andreas R\"uegg, and Jun Wen for useful discussions. We gratefully acknowledge financial support from ARO Grant No. W911NF-09-1-0527 and NSF Grant No. DMR-0955778. G.A.F acknowledges the hospitality of the Aspen Center for Physics under NSF Grant PHY-1066293 where part of this work was done. A. L. acknowledges partial support from the Alexander von Humboldt Foundation. For our exact diagonalizations, we have used the open source ALPS codes.
\vspace*{20pt}


\appendix
\section{Expressions of Bosonic Pairing and Hopping Terms \label{bosonH}}
In this Appendix we provide some details of Schwinger-boson men-field theory. The exchange interaction in terms of pairing and hopping of bosons in Eq.\eqref{exchange_decoupling} are described by the following expressions,
 \bea 
&&\hat{\chi}_{0,ij}=\frac{1}{2}(b^{\dag}_{i\uparrow}b_{j\uparrow}+b^{\dag}_{i\downarrow}b_{j\downarrow}),~~\hat{\Delta}_{0,ij}=\frac{1}{2}(b_{i\uparrow}b_{j\downarrow}-b_{i\downarrow}b_{j\uparrow})\nonumber\\
&&\hat{\chi}_{x,ij}=\frac{1}{2}(b^{\dag}_{i\uparrow}b_{j\downarrow}+b^{\dag}_{i\downarrow}b_{j\uparrow}),~~\hat{\Delta}_{x,ij}=\frac{1}{2}(b_{i\uparrow}b_{j\uparrow}-b_{i\downarrow}b_{j\downarrow})\nonumber\\
&&\hat{\chi}_{y,ij}=\frac{1}{2}(b^{\dag}_{i\uparrow}b_{j\downarrow}-b^{\dag}_{i\downarrow}b_{j\uparrow}),~~\hat{\Delta}_{y,ij}=\frac{1}{2}(b_{i\uparrow}b_{j\uparrow}+b_{i\downarrow}b_{j\downarrow})\nonumber\\
&&\hat{\chi}_{z,ij}=\frac{1}{2}(b^{\dag}_{i\uparrow}b_{j\uparrow}-b^{\dag}_{i\downarrow}b_{j\downarrow}),~~\hat{\Delta}_{z,ij}=\frac{1}{2}(b_{i\uparrow}b_{j\downarrow}+b_{i\downarrow}b_{j\uparrow}).\nonumber \\ \eea

\section{Hopping and Pairing Matrices \label{Hk}}
The $h_k$ and $\Delta_k$ in Eq.\eqref{H_k} are hopping and pairing matrices, respectively, whose elements are,
\begin{widetext}
\bea &&h^{11}_k=J_2\chi_{2}g_{2}-J_3\chi_{z}\cos(k\cdot a_2)+\lambda,~h^{12}_k=-J_3[\chi_{x}\cos k\cdot(a_1-a_2)+i\chi_y\sin(k\cdot a_1)],~h^{13}_k=\frac{1}{2}J_1\chi_{1}g_{1},~h^{14}_k=0,\nonumber\\&&h_k^{22}=J_2\chi_{2}g_{2}+J_3\chi_{z}\cos(k\cdot a_2)+\lambda,~h_k^{23}=0,~h^{24}_k=\frac{1}{2}J_1\chi_{1}g_{1},\nonumber\\
&&h^{33}_k=J_2\chi_{2}g_{2}-J_3\chi_{z}\cos(k\cdot a_2)+\lambda,~h^{34}_k=-J_3[\chi_{x}\cos k\cdot(a_1-a_2)+i\chi_y\sin(k\cdot a_1)],\nonumber\\
&&h^{44}_k=J_2\chi_{2}g_{2}+J_3\chi_{z}\cos(k\cdot a_2)+\lambda,\eea
\bea &&\Delta^{11}_k=J_3[\Delta_x\cos k\cdot(a_1-a_2)+\Delta_y\cos(k\cdot a_1)],~\Delta^{12}_k=iJ_2\Delta_{2}[\sin k\cdot(a_1-a_2)-\sin(k\cdot a_1)+\sin(k\cdot a_2)]+J_3\Delta_z\cos(k\cdot a_2),\nonumber \\ &&\Delta^{14}_k=\frac{1}{2}J_1\Delta_{1}g_1,~\Delta^{21}_k=iJ_2\Delta_{2}[-\sin k\cdot(a_1-a_2)+\sin(k\cdot a_1)-\sin(k\cdot a_2)]+J_3\Delta_z\cos(k\cdot a_2),\nonumber \\ &&\Delta^{22}_k=J_3[-\Delta_x\cos k\cdot(a_1-a_2)+\Delta_y\cos(k\cdot a_1)],~\Delta^{23}_k=-\frac{1}{2}J_1\Delta_{1}g_1,~\Delta^{32}_k=-\frac{1}{2}J_1\Delta_{1}g_1^{\ast},\nonumber\\ &&\Delta^{33}_k=J_3[\Delta_x\cos k\cdot(a_1-a_2)+\Delta_y\cos(k\cdot a_1)],~\Delta^{34}_k=iJ_2\Delta_{2}[\sin k\cdot(a_1-a_2)-\sin(k\cdot a_1)+\sin(k\cdot a_2)]+J_3\Delta_z\cos(k\cdot a_2),\nonumber\\ &&\Delta^{41}_k=\frac{1}{2}J_1\Delta_{1}g^{\ast}_1,~\Delta^{43}_k=iJ_2\Delta_{2}[-\sin k\cdot(a_1-a_2)+\sin(k\cdot a_1)-\sin(k\cdot a_2)]+J_3\Delta_z\cos(k\cdot a_2),\nonumber \\
&&\Delta^{44}_k=J_3[-\Delta_x\cos k\cdot(a_1-a_2)+\Delta_y\cos(k\cdot a_1)],~\Delta^{13}_k=\Delta^{24}_k=\Delta^{31}_k=\Delta^{42}_k=0,\eea 
where 
\bea g_1=1+e^{-ik\cdot a_1}+e^{-ik\cdot a_2},~~g_2=\cos(k\cdot a_1)+\cos(k\cdot a_2)+\cos k\cdot(a_1-a_2).\eea \end{widetext}

\section*{References}
%

\end{document}